\documentclass[twoside,twocolumn,9pt]{article}
\usepackage{extsizes}
\usepackage[super,sort&compress,comma]{natbib} 
\usepackage[version=3]{mhchem}
\usepackage[left=1.5cm, right=1.5cm, top=1.785cm, bottom=2.0cm]{geometry}
\usepackage{balance}
\usepackage{mathptmx}
\usepackage{sectsty}
\usepackage{graphicx} 
\usepackage{orcidlink}
\usepackage{booktabs}
\usepackage{multirow}

\usepackage{svg}
\usepackage{lastpage}
\usepackage[format=plain,justification=justified,singlelinecheck=false,font={stretch=1.125,small,sf},labelfont=bf,labelsep=space]{caption}
\usepackage{float}
\usepackage{fancyhdr}
\usepackage{fnpos}
\usepackage[english]{babel}
\addto{\captionsenglish}{%
  
}
\usepackage{array}
\usepackage{droidsans}
\usepackage{xspace}
\usepackage{charter}
\usepackage[T1]{fontenc}
\usepackage[usenames,dvipsnames]{xcolor}
\usepackage{setspace}
\usepackage[compact]{titlesec}
\usepackage{hyperref}

\usepackage{epstopdf}

\definecolor{cream}{RGB}{222,217,201}

\begin{document}

\pagestyle{fancy}
\thispagestyle{plain}
\fancypagestyle{plain}{
\renewcommand{\headrulewidth}{0pt}
}

\makeFNbottom
\makeatletter
\renewcommand\LARGE{\@setfontsize\LARGE{15pt}{17}}
\renewcommand\Large{\@setfontsize\Large{12pt}{14}}
\renewcommand\large{\@setfontsize\large{10pt}{12}}
\renewcommand\footnotesize{\@setfontsize\footnotesize{7pt}{10}}
\makeatother
\newcommand*{\MS}{MoS$_2$\xspace}
\newcommand*{\WS}{WS$_2$\xspace}
\newcommand*{\ME}{MoSe$_2$\xspace}
\newcommand*{\WE}{WSe$_2$\xspace}
\renewcommand{\thefootnote}{\fnsymbol{footnote}}
\renewcommand\footnoterule{\vspace*{1pt}%
\color{cream}\hrule width 3.5in height 0.4pt \color{black}\vspace*{5pt}} 
\setcounter{secnumdepth}{5}

\makeatletter 
\renewcommand\@biblabel[1]{#1}            
\renewcommand\@makefntext[1]%
{\noindent\makebox[0pt][r]{\@thefnmark\,}#1}
\makeatother 
\renewcommand{\figurename}{\small{Fig.}~}
\sectionfont{\sffamily\Large}
\subsectionfont{\normalsize}
\subsubsectionfont{\bf}
\setstretch{1.125} 
\setlength{\skip\footins}{0.8cm}
\setlength{\footnotesep}{0.25cm}
\setlength{\jot}{10pt}
\titlespacing*{\section}{0pt}{4pt}{4pt}
\titlespacing*{\subsection}{0pt}{15pt}{1pt}

\fancyfoot{}
\fancyfoot[LO,RE]{\vspace{-7.1pt}\includegraphics[height=9pt]{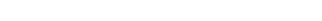}}
\fancyfoot[CO]{\vspace{-7.1pt}\hspace{13.2cm}\includegraphics{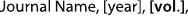}}
\fancyfoot[CE]{\vspace{-7.2pt}\hspace{-14.2cm}\includegraphics{head_foot/RF}}
\fancyfoot[RO]{\footnotesize{\sffamily{1--\pageref{LastPage} ~\textbar  \hspace{2pt}\thepage}}}
\fancyfoot[LE]{\footnotesize{\sffamily{\thepage~\textbar\hspace{3.45cm} 1--\pageref{LastPage}}}}
\fancyhead{}
\renewcommand{\headrulewidth}{0pt} 
\renewcommand{\footrulewidth}{0pt}
\setlength{\arrayrulewidth}{1pt}
\setlength{\columnsep}{6.5mm}
\setlength\bibsep{1pt}

\makeatletter 
\newlength{\figrulesep} 
\setlength{\figrulesep}{0.5\textfloatsep} 

\newcommand{\topfigrule}{\vspace*{-1pt}%
\noindent{\color{cream}\rule[-\figrulesep]{\columnwidth}{1.5pt}} }

\newcommand{\botfigrule}{\vspace*{-2pt}%
\noindent{\color{cream}\rule[\figrulesep]{\columnwidth}{1.5pt}} }

\newcommand{\dblfigrule}{\vspace*{-1pt}%
\noindent{\color{cream}\rule[-\figrulesep]{\textwidth}{1.5pt}} }

\makeatother

\twocolumn[
  \begin{@twocolumnfalse}
{\includegraphics[height=30pt]{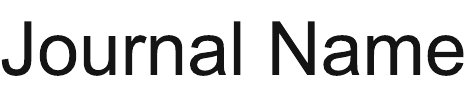}\hfill\raisebox{0pt}[0pt][0pt]{\includegraphics[height=55pt]{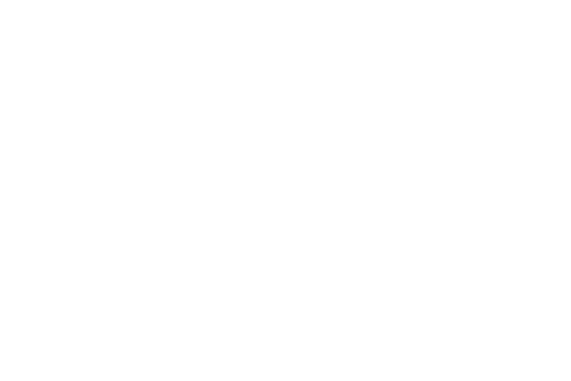}}\\[1ex]
\includegraphics[width=18.5cm]{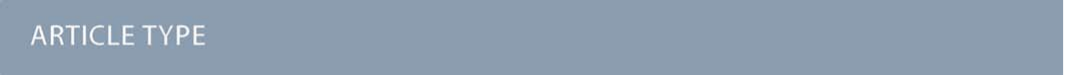}}\par
\vspace{1em}
\sffamily
\begin{tabular}{m{4.5cm} p{13.5cm} }

\includegraphics{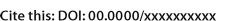} & \noindent\LARGE{\textbf{Janus MoSSe/WSSe Heterobilayers as Selective Photocatalysts for Water Splitting$\dag$}} \\
\vspace{0.3cm} & \vspace{0.3cm} \\

 & \noindent\large{Mostafa Torkashvand\orcidlink{0000-0002-5092-2711}\textit{$^{a}$}, Saeedeh Sarabadani Tafreshi\orcidlink{0000-0003-4130-437X}$^{\ast}$\textit{$^{a}$}, Caterina Cocchi\orcidlink{0000-0002-9243-9461}$^{\ast}$\textit{$^{b}$}, and  Surender Kumar\orcidlink{0009-0000-3072-5633}$^{\ast}$\textit{$^{b}$}   } \\

\includegraphics{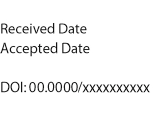} & \noindent\normalsize{Identifying materials that simultaneously straddle the water redox potentials and possess an intrinsic electric field is crucial for achieving high solar-to-hydrogen (STH) efficiency. Using state-of-the-art first-principles calculations, including a range-separated hybrid functional and spin-orbit coupling, we investigate MoXY/WXY (X, Y = S, Se) Janus bilayers for overall water splitting. {We find a critical competition between the metal-to-metal chemical potential difference and the intrinsic dipoles at the interface between the Janus monolayers.} We find that the Se–Se interfaced heterobilayer is intrinsically capable of driving water splitting, while its S–S counterpart can meet the redox requirements through pH modulation. For both configurations, a remarkable STH efficiency of 17.1\% is anticipated. {Furthermore, we predict a threshold of 1.0~eV for the built-in potential gradient to govern the transition from overall water splitting to band-edge pinning.} Compared to homobilayers, heterobilayers benefit from {the reciprocity between layer-specific dipoles and the Mo/W chemical potential difference, which promotes spatial separation and suppresses recombination, overall enhancing hydrogen production. Our results establish specific electronic descriptors for Janus heterostructures, providing a rational design rule for maximizing solar-driven hydrogen production in asymmetric two-dimensional materials.}
}

\end{tabular}
 \end{@twocolumnfalse} \vspace{0.6cm}]


\renewcommand*\rmdefault{bch}\normalfont\upshape
\rmfamily
\section*{}
\vspace{-1cm}


\footnotetext{\textit{$^{a}$~Department of Chemistry, Amirkabir University of Technology, 1591634311 Tehran, Iran; E-mail:  s.s.tafreshi@aut.ac.ir}}
\footnotetext{\textit{$^{b}$~Institut für Festk\"orpertheorie und -Optik, Friedrich-Schiller-Universit\"at Jena, 07743 Jena, Germany, E-mail:  caterina.cocchi@uni-jena.de, surendermohinder@gmail.com }}
\footnotetext{\dag~Supplementary Information available: [details of any supplementary information available should be included here]. See DOI: 00.0000/00000000.}



\section*{Introduction}
  Photocatalytic water splitting is regarded as a promising carbon-neutral route for hydrogen production through the direct conversion of solar energy into chemical fuel.\cite{Yin2021,Maeda2010,Nishioka2023,Zou2001,Li2014} Given the global abundance of solar energy and water, hydrogen generation via photocatalytic water splitting is highly sustainable, environmentally friendly, and potentially cost-effective.\cite{Nishioka2023,Acar2016,Zou2001}  
  The underlying mechanism for photocatalytic water splitting begins with the excitation of the photocatalyst, which generates electron–hole pairs upon photo-absorption. The generated charge carriers play a central role in redox reactions, and their rapid separation and transport are essential to suppress charge recombination promoted by Coulomb attraction.\cite{Zou2001,Liu2023} 
  
  An effective photocatalyst for water splitting should have a conduction band minimum (CBM) higher than the reduction potential (-4.44~eV for $H^+/H_2$ relative to vacuum) and the valence band maximum (VBM) lower than the oxidation potential (-5.67~eV for $O_2/H_2O$).\cite{Zou2001,Ju2020,Maeda2010,Nishioka2023,Liu2023} Moreover, a band gap exceeding the thermodynamic threshold of 1.23~eV is necessary for maximizing the solar-to-hydrogen (STH) efficiency.\cite{Fu2018,ye2023,Rawat2020,Bao22025,Maeda2010,Nishioka2023} While redox ability requires a large band gap, visible light-harvesting is optimized by a fundamental gap energy falling in the visible region. These two conditions, which are crucial for high efficiency,\cite{Fu2018}  are challenging to achieve simultaneously and demand the design of novel materials. 

In the last decade, two-dimensional (2D) transition metal dichalcogenides (TMDs) have emerged as key materials in photocatalytic research.\cite{Ju2020,ye2023,Wei2019,Ji2018,Guan2018,Acar2016,Rawat2020,Bao22025,Ju2020a,Ma2018} These systems are uniquely suited for solar-driven water splitting due to their robust chemical stability, band gaps within the visible spectrum, and exceptional structural tunability.\cite{Ma2018,Ju2020,ye2023,Rawat2020,Bao22025,XU2022,Guan2018,Wei2019} TMDs offer a large compositional and structural versatility, including from pristine monolayers to asymmetric Janus structures and complex van der Waals (vdW) heterostructures.\cite{Ju2020,Rawat2020,Ma2018,ye2023,Bao22025,XU2022,Guan2018,Fu2018,xia2018,Wei2019} This diversity allows a precise control over electronic properties through thickness modulation and tailored stacking,\cite{Ma2018,Ju2020,ye2023,Rawat2020,Bao22025,XU2022,Guan2018,Wei2019} maximizing light harvesting and enhancing charge carrier separation, which are the fundamental requirements for high-performance photocatalysis.

Due to their broken out-of-plane structural symmetry, Janus TMDs have recently emerged as a particularly promising material platform for photocatalysis.\cite{Ma2018,Ju2020,ye2023,Rawat2020,Bao22025,XU2022,Guan2018,Wei2019} By substituting one layer of chalcogen atoms with another species in the same group (e.g., S $\rightarrow$ Se)\cite{Zhang2017,Lu2017}, a static dipole moment is generated perpendicular to the basal plane, giving rise to a permanent internal polarization that is absent in conventional, symmetric monolayers. \cite{Ma2018,Ju2020,ye2023,Rawat2020,Bao22025,XU2022,Guan2018,Fu2018,xia2018} This built-in electric field can act as an intrinsic driving force for water splittingby producing a continuous potential gradient that effectively separates photo-generated charge carriers.  While monolayers and homo-metal Janus bilayers have been largely investigated to achieve tunable band alignment and enhanced carrier lifetimes,\cite{Ma2018,Ju2020,ye2023,Rawat2020,Bao22025,XU2022,Guan2018,Wei2019} 

{TMD Janus heterobilayers represent a fundamentally distinct class of vdW structures, characterized by a highly tunable chemical and electrostatic landscape given by the numerous combinations between metallic and chalcogen species. Including different transition metals,\cite{alfurhud2024,xia2018}, these systems introduce a synergy between layer-specific Janus dipoles and an interfacial chemical potential difference. This reciprocity further modulates the built-in electric field,\cite{Fu2018,xia2018} not only promoting selective hydrogen and oxygen evolution reactions (HER and OER, respectively) on distinct surfaces but also dictating the migration pathways of photogenerated carriers.\cite{Liu2023}  Consequently, strategic stacking can effectively suppress interfacial charge recombination and maximize carrier utilization, ultimately leading to superior STH conversion yields.}


Here, we investigate from first-principles the electronic structure and photocatalytic ability of TMD Janus heterobilayers with the chemical formula MoXY/WXY (where X, Y $\in$ {S, Se}). By exploring four configurations with AB-stacking and comparing them against their homo-metal counterparts, we {rationalize the competition between layer-specific dipoles and the metal-to-metal chemical potential difference acting on the built-in electric field, to identify the thermodynamic window for water splitting.} Our analysis of band offsets and pH-dependent redox potentials identifies two specific configurations that are capable of spontaneous overall water splitting. The stacking-dependent work function dictates the optimal operational environment: the bilayer with the highest work function is optimized for acidic conditions ($\text{pH} = 0$), while the lowest work function imposes alkaline conditions ($\text{pH} = 12.5$) to align with the Nernstian-shifted redox levels. For these systems, we obtain a remarkable STH efficiency of 17.14\%. {By defining the electronic prerequisites and the spatial distribution of photogenerated carriers, this work establishes rational design rules for maximizing the performance of Janus-based photocatalysts.}


\section*{Computational details}
We performed density-functional theory (DFT) calculations with the projector augmented wave (PAW) method for electron-ion interaction, as implemented in the Vienna Ab initio Simulation Package (VASP).~\cite{Hohenberg1964,Kresse1996,Kresse1999,blochl1994} The exchange-correlation functional is treated within the generalized gradient approximation of Perdew, Burke, and Ernzerhof (PBE)\cite{Perdew1996} during relaxation, while the Heyd-Scuseria-Ernzerhof (HSE06) screened hybrid functional~\cite{Heyd2003,Heyd2006} is adopted to compute the electronic and photocatalytic properties.  Structural optimization is conducted until the total energy is below $10^{-6}$ eV and the Hellmann-Feynman forces are less than 0.01 eV/\AA{} per atom. To account for long-range dispersion forces, the DFT-D3 semi-empirical correction is applied.\cite{Grimme2010,Grimme2011} A plane-wave basis set with a kinetic energy cutoff of 550~eV and a $\Gamma$-centered $15 \times 15 \times 1$ \textbf{k}-point grid is used for all systems. In the HSE06 calculations, the total-energy convergence threshold is reduced to $10^{-5}$~eV.  A vacuum layer exceeding 20~\AA{} is included along the \textit{z}-direction to eliminate spurious interactions between periodic images.  Given the broken inversion symmetry and intrinsic polarity of the Janus layers, a dipole correction is included to ensure an accurate alignment to the vacuum level.  Spin-orbit coupling (SOC) is included in the electronic property calculations. The post-processing analysis is  done with the VASPKIT~\cite{WANG2021} tool and in-house-produced \texttt{Python} scripts.
\begin{table*}[h] 
\centering
\caption{Structural properties of hetero metal bilayers: Lattice constants (a), metal-metal distance (d), various metal-chalcogenide bond-lengths and binding energies (E$_b$). 
}
\label{tab:hetero_properties}
\renewcommand{\arraystretch}{1.2}
\begin{tabular}{l c c c c c c c c c }
\hline
Configuration & $a$ (\AA) & $d$ (\AA) & $d_{\rm Mo-S}$ (\AA{}) & $d_{\rm Mo-Se}$ (\AA{}) &$d_{\rm W-S}$ (\AA{}) & $d_{\rm W-Se}$ (\AA{})  & E$_{b}$ (eV)\\
\hline
\multirow{4}{*}{}
SMoSe$\mid$SWSe& 3.23 & 6.62 & 2.41 & 2.53 & 2.42 &2.53& -2.23 \\
SeMoS$\mid$SWSe& 3.23 & 6.40 & 2.41 & 2.53 & 2.42 &2.53&-2.21\\
SMoSe$\mid$SeWS& 3.23 & 6.89 & 2.41 & 2.53 & 2.42 &2.53&-2.24 \\
SeMoS$\mid$SeWS& 3.23 & 6.62 & 2.41 & 2.53 & 2.42 &2.53&-2.22 \\
\hline

\end{tabular}
\end{table*}
\section*{Results and discussion}

\begin{figure}
    \centering
    \includegraphics[width=\linewidth]{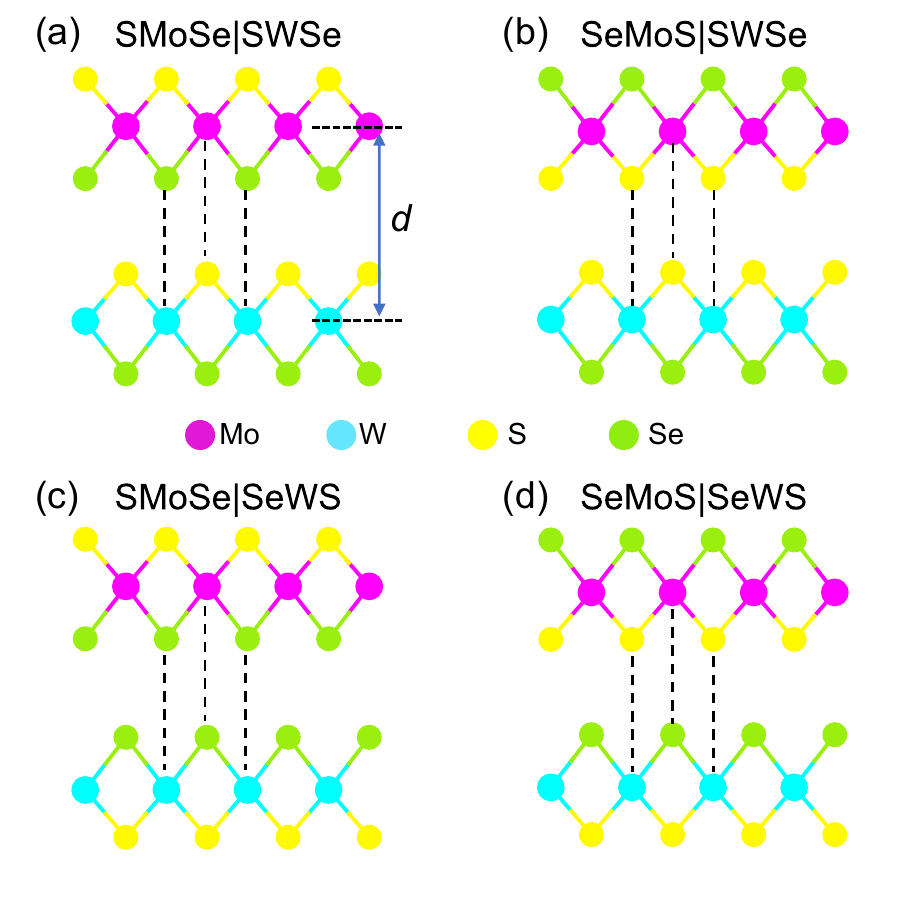}
    \caption{Side views of the optimized AB-stacked MoSSe/WSSe heterobilayers in the 2H phase: (a) SMoSe|SWSe, (b) SeMoS|SWSe, (c) SMoSe|SeWS, and (d) SeMoS|SeWS. The blue arrow in panel (a) indicates the metal–metal distance $d$. The gray dashed bars guide the eyes to recognize the stacking configuration. }
    \label{fig:str}
\end{figure} 

\subsection*{Structural properties}
We investigate four AB-stacked configurations of TMD Janus heterobilayers composed of MoSSe and WSSe.\cite{alfurhud2024} These systems are categorized by their chalcogen interfaces,\cite{alfurhud2024,Nielsen2025} with either the same species (S|S or Se|Se) or different species (S|Se or Se|S) facing each other (Fig.~\ref{fig:str}).  
In the considered AB-stacking, the metal atom in one Janus monolayer aligns vertically with a chalcogen atom in the other monolayer. To avoid the lattice mismatch that naturally emerge from the difference in the in-plane lattice constants of the two monolayers, ranging between $3.16$~\AA{}\cite{Wildervanck1964, Coehoorn1987,Dungey1998,Yun2012} to $3.29$~\AA{}\cite{Nielsen2025b,Ju2020,Chen2017,Yun2012} in the Mo- and W-based sheets, respectively, we assign to all configurations an initial lattice constant of $a=3.29$~\AA{}. Upon structural optimization, all heterobilayers maintain the same value of $a=3.23$~\AA{}, which, however, substantially reduces compared to the input value, effectively averaging the lattice constants of the single constituents (Table~\ref{tab:hetero_properties}). This results in inherent strain that, combined with the broken mirror symmetry of the Janus monolayers, uniquely defines the structural properties of these heterostructures.

The metal-chalcogen bond lengths are nearly identical in all considered systems, in agreement with previous theoretical and experimental reports.~\cite{Nielsen2025,xia2018,alfurhud2024,Guo2020} In contrast, the interlayer distance $d$, defined as the metal–metal separation along the out-of-plane direction, shows a strong dependence on the chalcogen interface (Table~\ref{tab:hetero_properties}). Specifically, Se|Se interfaced bilayers display the largest interlayer distance $d=6.89$~\AA{}, whereas those with an S|S interface show the smallest separation ($d=$6.40~\AA{}). For mixed S|Se interfaces, the interlayer distance adopts an intermediate value of 6.62~\AA{}, approximately equal to the average of the other two.\cite{Guo2020,Nielsen2025} 

The relative stability of the heterostructures is estimated by their binding energy, defined as
\begin{equation}
 E_{\rm b} = E_{\text{hetero}} - E_{\rm MoSSe} - E_{\rm WSSe},
 \label{eq:E_b}
\end{equation}
 where $E_{\text{hetero}}$, $E_{\rm MoSSe}$, and $E_{\rm WSSe}$ are the DFT total energies of the heterobilayers and their isolated monolayer constituents, respectively. All calculated binding energies are negative [Table\ref{tab:hetero_properties}, for homo-metals see the Electronic Supplementary Information (ESI), Table~S1], confirming energetic stability, and range between -2.24~eV in SMoSe$\mid$SeWS to -2.21~eV in SeMoS$\mid$SWSe.
 {These values demonstrate superior structural stability compared to standard TMD heterobilayers.\cite{amin2015heterostructures,liu2023role} Given the established dynamical stability of Janus monolayers,\cite{van2020first} it is expected that these heterostructures will remain robustly bound under ambient conditions.}

\subsection*{Electronic properties}
\begin{figure*}
    \centering
    \includegraphics[width=0.45\textwidth]{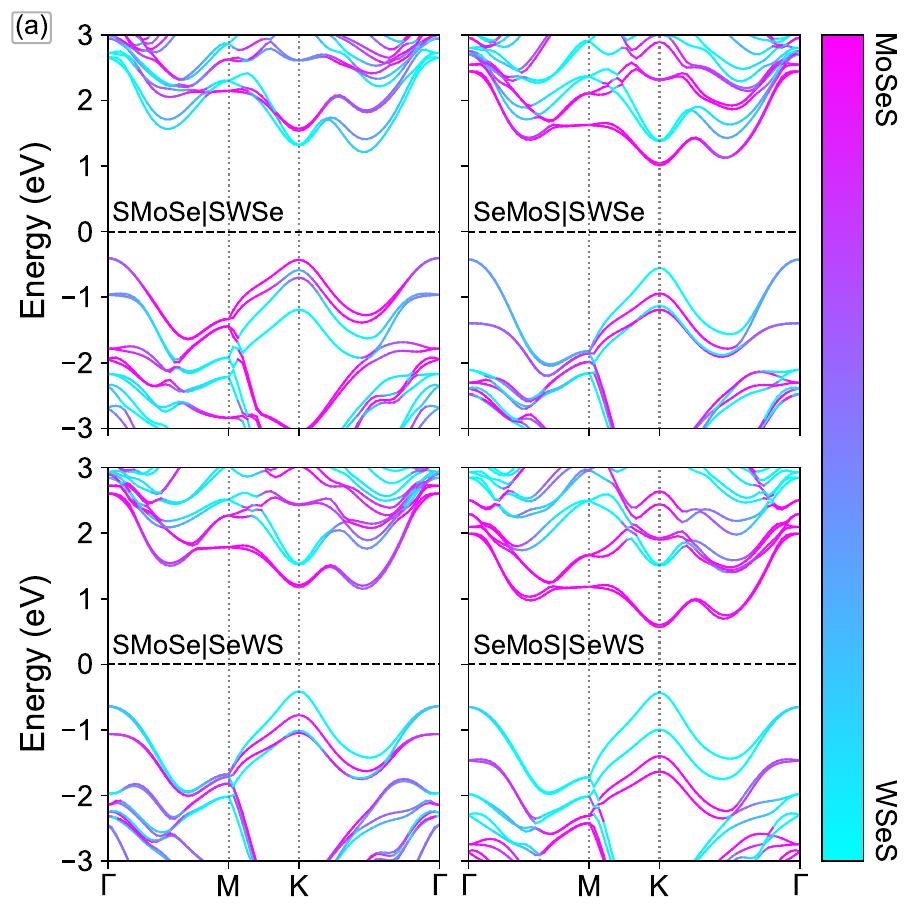}
    \includegraphics[width=0.45\textwidth]{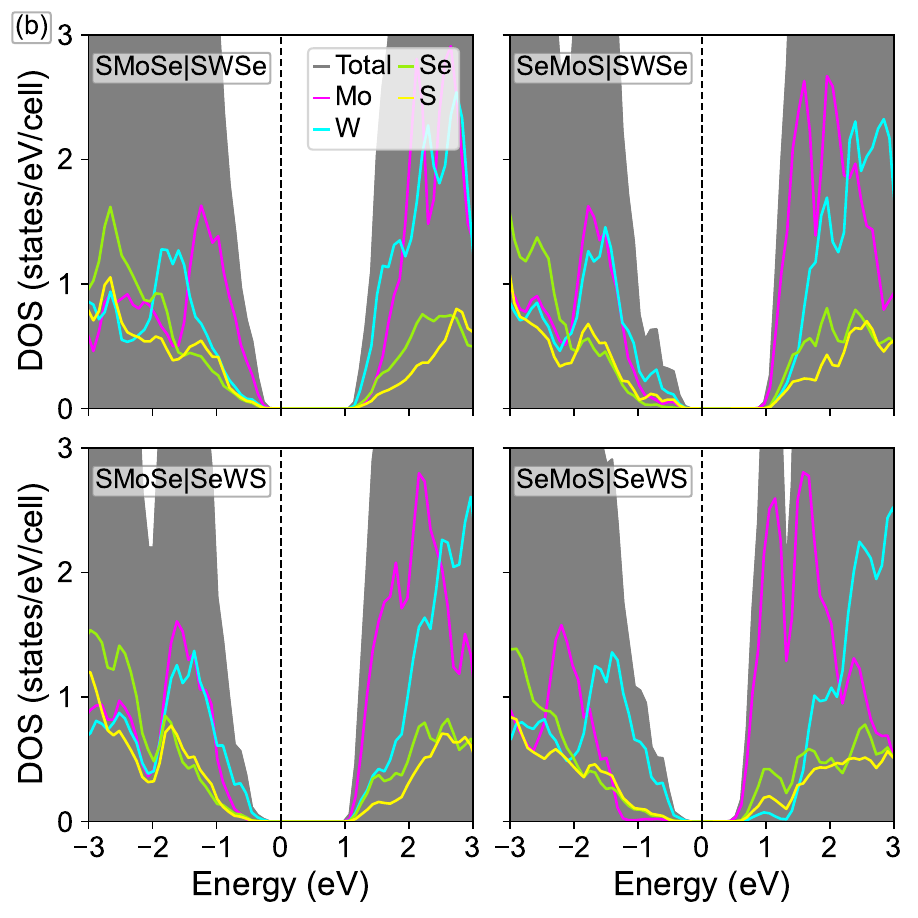}
    \caption{(a) Layer-projected electronic band structures of the considered Janus heterobilayers calculated using the HSE06 hybrid functional with spin-orbit coupling.   (b) Total (gray area) and atom-projected density of states. In all graphs, the Fermi level is set to 0~eV.  }
    \label{fig:layer_band}
    \end{figure*}

The considered Janus heterobilayers show distinct electronic band structures (Fig.~\ref{fig:layer_band}(a)). Except for the SeMoS$\mid$SeWS heterostructure, which exhibits a direct band gap at the high-symmetry point K, the other systems are characterized by an indirect band gap. In the SMoSe$\mid$SWSe bilayer, featuring a fundamental gap of 1.61~eV (Table~\ref{tab:electronic_properties}), the VBM at $\Gamma$ is primarily localized on the MoSSe layer, while the CBM, exhibiting signatures of interlayer hybridization via mixed orbital contributions, appears between K and $\Gamma$, at the so-called Q-valley.\cite{Nielsen2025}
Conversely, in the SeMoS$\mid$SWSe bilayer, the VBM remains at $\Gamma$ while the CBM appears at the K-point, leading to an indirect gap of 1.43~eV. Here, the layer contributions are reversed compared to SMoSe$\mid$SWSe, with the VBM arising from WSSe and the CBM from MoSSe. In SMoSe$\mid$SeWS, the uppermost valence band at K is above the $\Gamma$-valley, while the lowest CB at Q remains below the K-point, leading to an indirect gap of 1.56~eV with VBM and CBM localized on the WSSe and MoSSe monolayers, respectively. Finally, the SeMoS$\mid$SeWS bilayer has a direct band gap of 1.00~eV at the K-valley. Remarkably, in all these AB-stacked Janus heterostructures, the frontier bands remain mostly localized on a specific monolayer, indicating suppressed hybridization.

\begin{table*}[h]
\centering
\caption{ Calculated band gaps from HSE06 (E$_{g}^{HSE06}$) with the direct one obtained for SeMoS$\mid$SeWS bolded, dipole moments ($\mu$), electrostatic potential difference ($\Delta \Phi$) within the heterobilayers, and work function (W).  
}
\label{tab:electronic_properties}
\renewcommand{\arraystretch}{1.2}
\begin{tabular}{l c c c c c c}
\hline
Configuration &$E_g^{\text{HSE06}}$ (eV) &   $\mu$ (D) & $\Delta \Phi$ (eV) &$W$ (eV) \\
\hline
\multirow{4}{*}{}
  SMoSe$\mid$SWSe & 1.61& 0.346 & 1.32& 5.72\\
  SeMoS$\mid$SWSe &1.43 & 0.002 & 0.18 & 4.60\\
  SMoSe$\mid$SeWS &1.56 &  0.017& 0.18 &5.51 \\
 SeMoS$\mid$SeWS &\textbf{1.00} & $-0.321$ & 1.09 & 5.30 \\
\hline
\end{tabular}
\end{table*} 

The analysis of the projected density of states (PDOS) (Fig.~\ref{fig:layer_band}(b)) reveals the atomic contributions to the electronic structure of the considered heterobilayers. As expected,\cite{Chang2013,Guo2020,Patel2022,Wei2019} the band edges are dominated by contributions from the transition-metal species, Mo and W, which strongly influence the overall shape of the PDOS and allow retrieving the trends identified in Fig.~\ref{fig:layer_band}(a), particularly the type-II level alignment. While this characteristic is common to all heterobilayers, the spatial localization of the VBM and CBM is sensitive to the composition. In SMoSe|SWSe, the VBM is localized on the Mo-containing monolayer and the CBM on the W-based one, whereas the opposite is true in all other systems.

\begin{figure}
    \centering
    \includegraphics[width=\linewidth]{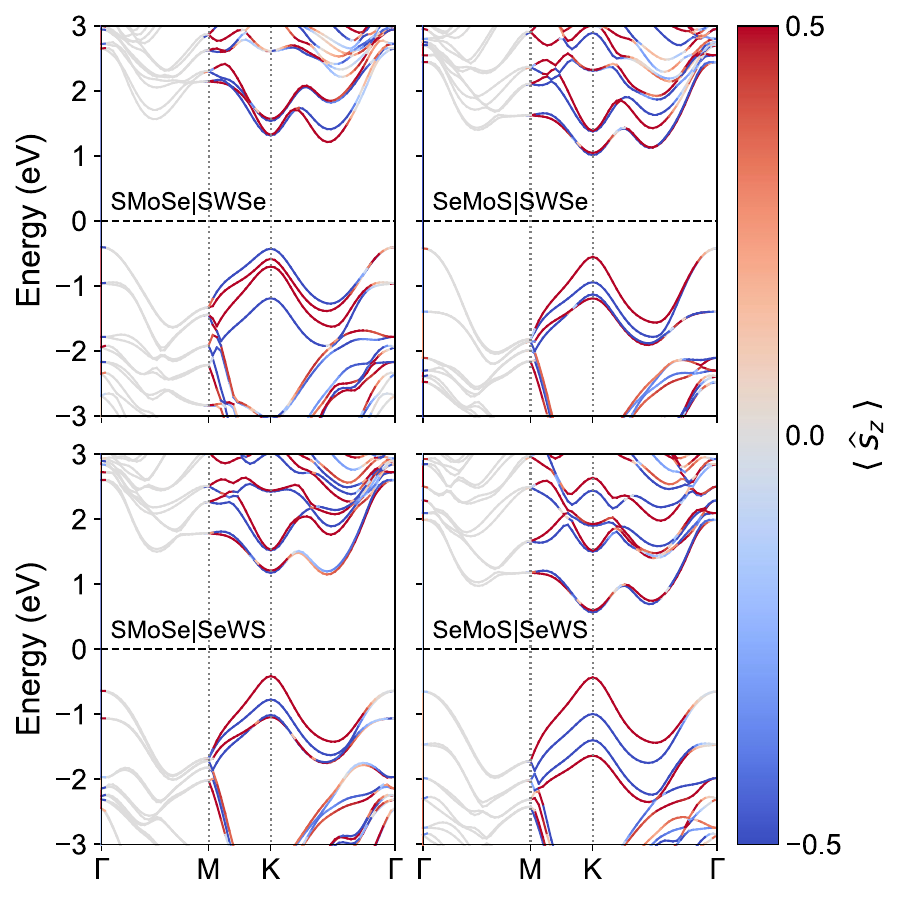}
    \caption{Spin-projected electronic band structures ($s_z$ component) of the Janus heterobilayer configurations, calculated at the HSE06+SOC level of theory. The bands are color-coded based on their $s_z$ spin expectation values to visualize spin-splitting and potential Rashba-like features. In all panels, the Fermi level ($E_F$) is shifted to 0 eV, indicated by horizontal dashed line.}
    \label{fig:spin_band}
\end{figure}

We deepen the electronic-structure analysis by inspecting in detail the effects of spin-polarization and SOC, highlighted in the spin-projected band structures presented in Fig.~\ref{fig:spin_band}. In the SMoSe|SWSe bilayer, the CBM at the Q-valley shows a significant spin splitting, whereas the VBM splitting at the K-point remains small. This is consistent with the characteristics of the constituents: the CBM is primarily dominated by the heavy W atoms in the WSSe layer, which possess a larger atomic SOC constant, while the VBM is stems from Mo states in the MoSSe monolayer. A similar behavior is observed across other configurations, where bands associated with the WSSe monolayer shows larger spin-splitting values ($\sim 600$~meV) compared to those localized in MoSSe. The large spin-orbit splitting affecting the WSSe monolayers render the band-gap nature of SeMoS$\mid$SeWS particularly sensitive to relativistic effects: without SOC, this system features an indirect band gap, as predicted in previous reports~\cite{alfurhud2024}. In fact, ignoring SOC, the VBM at the K-valley resides slightly lower than the $\Gamma$-point, while when SOC is included, the VBM shifts to the zone center, although the energies at these two high-symmetry points remain very close. Another important characteristic of these Janus heterobilayers emerging from Fig.~\ref{fig:spin_band} is the spin-polarization of the band edges. While in SeMoS|SWSe, SMoSe|SeWS, and SeMoS|SeWS the highest-occupied and lowest-unoccupied states at the K-valley have parallel spins, in SMoSe|SWSe, both the CBM at Q and the lowest-energy conduction state at K have spin-up while the uppermost valence band has spin-down. This feature is expected to impact the optical absorption of these heterobilayers. An in-depth analysis on this behavior will be the subject of up-coming work.

The presence of different chalcogen species in the Janus structures creates a built-in electric field within each monolayer.\cite{Li2017,Ji2018,Patel2022,Wei2019} When these monolayers are combined to form a heterobilayers, the intrinsic dipoles of the constituents generate an interfacial electrostatic potential which is further modulated by the mutual orientation of the monolayers (Fig.~\ref{fig:pot}). 
The varying depth of the electrostatic potential arises precisely from the different electronegativity and size of the atoms, being therefore maximized in WSSe. The orientation of the generated field is determined by the chalcogen arrangement, expectedly pointing from Se to S, namely from the more positively to the more negatively charged regions.

In the SMoSe$\mid$SWSe and SeMoS$\mid$SeWS configurations, the intrinsic dipole moments of the monolayers point to the same direction, resulting in a strong additive effect. This leads to substantial potential differences of 1.32~eV and 1.09~eV, respectively, and to total dipole moments up to 0.346~D (Table~\ref{tab:electronic_properties}). For the remaining two configurations, the monolayers are oriented with opposing internal fields that nearly cancel each other out, resulting in a negligible net electric field and negligible dipole moments equal or below 0.02~D. This dramatic difference in the internal field strength directly influences the band alignment and the spatial separation of charge carriers across the interface.

The work functions of the heterobilayers, computed from the difference between the vacuum level and the Fermi energy (Table~\ref{tab:electronic_properties}), provide another relevant indicator regarding the potential performance of these systems in photocatalysis. The SeMoS$\mid$SWSe bilayer exhibits the lowest work function (4.60~eV), whereas SMoSe$\mid$SWSe shows the highest value (5.72~eV). This large variation originates from interface-dependent charge redistribution and the resulting dipole-induced shift of the vacuum level. Such work-function differences directly determine the absolute band-edge alignment with aqueous redox potentials, thereby governing carrier transfer at the semiconductor–electrolyte interface and, consequently, the efficiency of photocatalytic water splitting, as discussed below.

\subsection*{Photocatalytic properties}
\begin{figure*}[h]
    \centering
    \includegraphics[width=\textwidth]{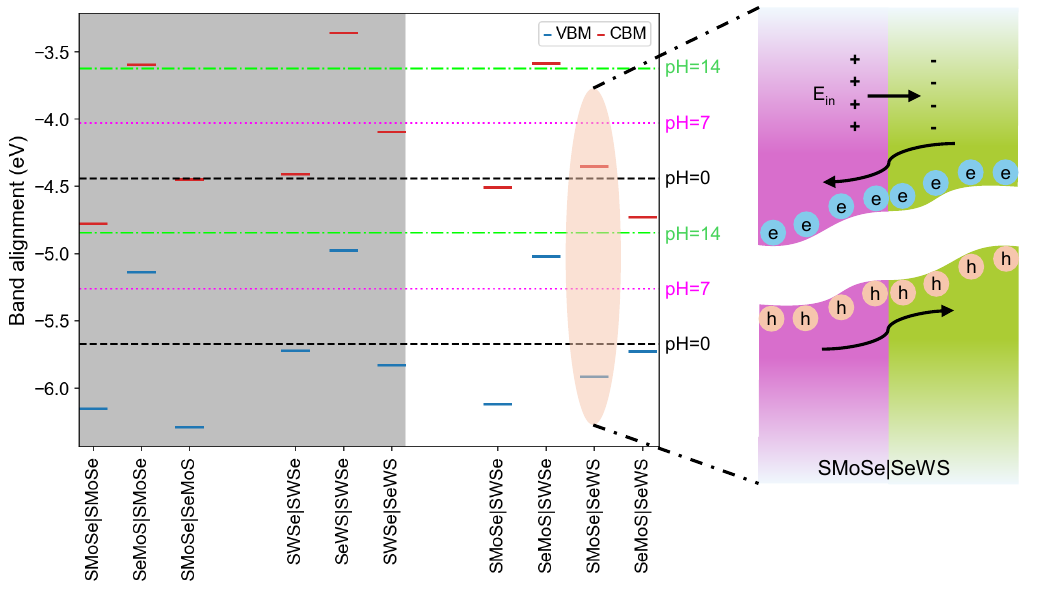}
    \caption{Band alignment of homo-metal (gray background) and hetero-metal Janus bilayers relative to the water redox potentials referenced to the vacuum level set to 0~eV. The scheme on the right illustrates the band offsets and the direction of the built-in electric field in the SeMoS$\mid$SeWS heterobilayer.}
    \label{fig:offset}
\end{figure*}

With the detailed knowledge of the electronic structure of the TMD Janus heterobilayers, we now evaluate their performance in photocatalytic water splitting by comparing their band edges with the water redox potentials. We recall that the overall water-splitting mechanism includes both the OER~\cite{Fu2018,xia2018,Guan2018,Ju2020}, $4h^+ + 2H_2O \rightarrow O_2 + 4H^+$ and the HER, 
$4e^- + 4H_2O \rightarrow 2H_2 + 4OH^-$.
In this process, photoexcited electrons are transferred from the CBM of the catalyst to drive the reduction process, while photogenerated holes at the VBM facilitate the oxidation process. Consequently, the thermodynamic feasibility of these reactions heavily depends on the absolute values of the CBM and VBM relative to the standard reduction potential of $H^+/H_2$ at 4.44~eV and the oxidation potential of $H_2O/O_2$ at -5.67~eV (pH$=0$). Beyond the requirement that the band edges straddle the redox potentials, the kinetic overpotentials, namely the energy difference between the band edges and the redox levels, should be large enough to overcome the activation barriers of each half-reaction.
\begin{figure}
    \centering
    \includegraphics[width=\linewidth]{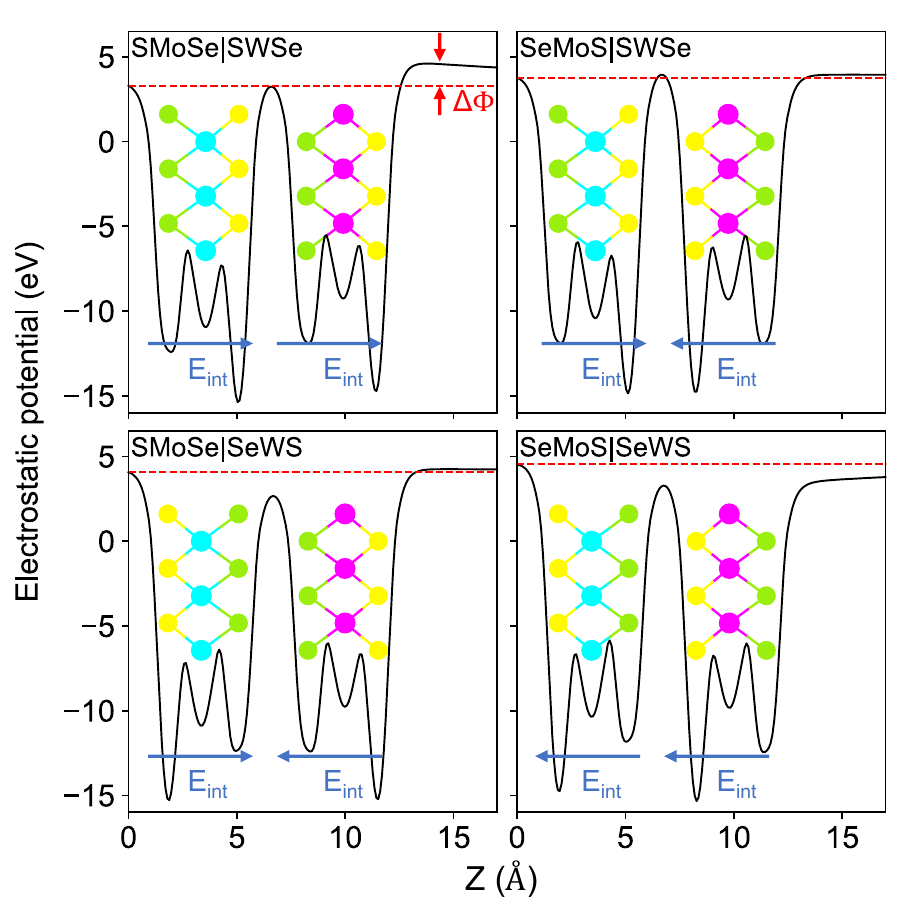}
    \caption{ One-dimensional profile of the planar-averaged Hartree potential for various heterobilayers. Red arrows denote the potential difference ($\Delta \Phi$) between the two surfaces of the bilayer, while blue arrows indicate the intrinsic internal electric field within each Janus monolayer.}
    \label{fig:pot}
\end{figure}
%
An overview of the frontier energy levels of the considered Janus heterobilayers relative to the redox potential, compared for reference with the same results computed for the homobilayers, is reported in Fig.~\ref{fig:offset}. Similar to the homobilayers featuring a Se-Se interface (SMoSe$\mid$SeMoS and SWSe$\mid$SeWS), which are the only ones fulfilling the requirement for overall catalytic water splitting at pH$=0$,\cite{xia2018,Guan2018,Yin2018a,Wei2019,Ji2018} among the heterobilayers, only SMoSe$\mid$SeWS exhibits the desired band alignment. In this system, the CBM is 80~meV above the $H_2$ reduction level, and the VBM is 250~meV below the oxygen reduction threshold, demonstrating the catalytic robustness of this heterostructure. 

The remaining three heterobilayers fulfill only one condition at pH$=0$. In SMoSe$\mid$SWSe, the CBM lies approximately 100~meV below the $H_2$ reduction level, rendering it incapable of hydrogen evolution, although its VBM is positioned deep enough to effectively oxidize $O_2$. 
Conversely, in SeMoS$\mid$SWSe, the VBM is over 500~meV above the $O_2$ redox level, making it unsuitable for oxygen evolution, even though the CBM is well-positioned for $H_2$ reduction. The SeMoS$\mid$SeWS configuration lacks the necessary band gap threshold of 1.23~eV. Moreover, while its VBM lies slightly below the $O_2$ redox level, its CBM is significantly below the $H_2$ level, preventing the HER. 

The results presented in Fig.~\ref{fig:offset} suggest that the Se-Se interface is a crucial requirement for Janus bilayers to operate as effective photocatalysts at pH$=0$.
However, the different chemical potentials (or electronegativities) of the metal atoms in the heterobilayers lead to the formation of a permanent dipole moment, which represents a strategic advantage compared to the homobilayers, where the internal fields from each layer cancel out completely. As shown schematically in Fig.~\ref{fig:offset} (left), the MoSSe monolayer has a slightly more positive charge than WSSe, as expected from the deeper electrostatic potential of W compared to Mo (compare Fig.~\ref{fig:pot}). This imbalance creates an internal electric field directed from the MoSSe toward the WSSe layer. Because the CBM of MoSSe is lower than that of WSSe, this built-in electric field facilitates the migration of electrons from WSSe to MoSSe. Concomitantly, the higher VBM in WSSe compared to MoSSe allows the former to accept holes from the latter, a process that would be inhibited in the absence of an electric field. 

Based on these considerations, the intrinsic dipole moment of the Janus heterobilayers is expected to promote the spatial separation of photogenerated electrons and holes into different layers. By physically isolating charge carriers of opposite sign, the exciton recombination rate is likely reduced, allowing efficient water splitting, particularly in the  SMoSe$\mid$SeWS.
\subsubsection*{Photocatalytic Performance Modulation}
The efficiency of photocatalytic water splitting is intimately linked to the pH of the aqueous environment, which controls the positions of the water redox potentials. 
The pH-dependent redox potentials at room temperature are governed by the Nernst equation\cite{Bolts1976,Pham2014,Xin2015,xia2018}: $E_{pH} = E_{pH=0} - 0.059 \cdot pH$ where $E_{pH}$ is the redox potential at a specific acidity or alkalinity level. As the pH value increases from 0 to 14, the redox potentials for both HER and OER shift upward in energy. 

To account for pH variations depending on the catalytic environment, we reassess the alignment of the frontier levels of the considered heterobilayers (and homobilayers for reference) with respect to the modified values of the HER and OER potentials at pH$=7$ and pH$=14$ (Figure~\ref{fig:offset}).
At pH$=7$ (neutral environment), none of the considered systems fulfill the conditions for HER and OER simultaneously. On the other hand, at extreme alkaline conditions (pH$=14$), both the SeWS|SeWS homobilayer and the SeMoS$\mid$SWSe heterobilayer, which are inactive at lower pH values, become capable of overall water splitting in the range $11.5\leq$pH$\leq14$, as their band edges successfully straddle the shifted redox levels. 

The overall water splitting ability of a catalyst is quantified by its STH efficiency ($\eta_{\rm STH}$),\cite{Fu2018,xia2018,Ju2020} defined as the product of the light absorption efficiency ($\eta_{\rm abs}$) and the carrier utilization efficiency ($\eta_{\rm cu}$)~\cite{Fu2018,xia2018,Guan2018,Ju2020}:
\begin{equation}
    \eta_{\rm STH}=\eta_{\rm abs}\cdot\eta_{\rm cu}\, .
\end{equation}
The light absorption efficiency represents the fraction of the total incident solar power that the semiconductor can potentially absorb, determined by its band gap and the standard solar flux spectrum, $P(\hbar \omega)$:
\begin{equation}
\eta_{\rm abs} =  \frac{\int_{E_{g} }^{\infty} P(\hbar \omega) d(\hbar \omega)}{\int_{0}^{\infty} P(\hbar \omega) d(\hbar \omega)} \, .
\end{equation}
The carrier utilization efficiency is the ratio of the generated chemical energy to the total absorbed solar power, calculated as:
\begin{equation}
\eta_{\rm cu} =  \frac{ \Delta G \int_{E }^{\infty} \frac{P(\hbar \omega)}{\hbar \omega} d(\hbar \omega)}{\int_{E_{g}}^{\infty} P(\hbar \omega) d(\hbar \omega)},
\end{equation} 
where $\Delta G = $1.23~eV represents the variation of the Gibbs free energy for water splitting. The effective energy required to drive the reaction accounts for the catalytic overpotentials of HER ($\chi(\mathrm{H}_2)$) and OER ($\chi(\mathrm{O}_2)$). The results obtained for the SMoSe|SeWS bilayer at pH$=0$ and the SeMoS|SWSe bilayer at pH$=12.5$ are summarized in Table~\ref{tab:STH}.

To ensure sufficient driving force, the overpotentials are typically compared against empirical benchmarks (0.2~eV for $\text{H}_2$ and 0.6~eV for $\text{O}_2$), leading to the following conditional form for $E$~\cite{Fu2018,xia2018,Guan2018,Ju2020}: 
\begin{equation}
    E =
\begin{cases}
E_g,
& \chi(\mathrm{H}_2) \ge 0.2,\; \chi(\mathrm{O}_2) \ge 0.6 \\[4pt]

E_g + (0.2 - \chi(\mathrm{H}_2)),
& \chi(\mathrm{H}_2) < 0.2,\; \chi(\mathrm{O}_2) \ge 0.6 \\[4pt]

E_g + (0.6 - \chi(\mathrm{O}_2)),
& \chi(\mathrm{H}_2) \ge 0.2,\; \chi(\mathrm{O}_2) < 0.6 \\[4pt]

E_g + (0.2 - \chi(\mathrm{H}_2)) + (0.6 - \chi(\mathrm{O}_2)),
& \chi(\mathrm{H}_2) < 0.2,\; \chi(\mathrm{O}_2) < 0.6
\end{cases}
\end{equation}
For both configurations, the HER overpotentials are impressively low ($\chi(\text{H}_2)=$0.08~eV), while the OER ones remain well below the 0.6 eV. These minimal kinetic barriers ensure that a significant portion of the absorbed photon energy is effectively utilized to drive the redox reactions rather than being lost to overpotential requirements. 
Remarkably, both configurations yield an identical $\eta_{\text{STH}}=17.14\%$, despite a difference in their band gaps of approximately 130~meV. 

\begin{table}[!ht]
\centering
\caption{OER and HER overpotential ($\chi(\mathrm{O}_2)$ and $\chi(\mathrm{H}_2)$, in eV), absorption efficiency ($\eta_{\rm abs}$), carrier utilization efficiency ($\eta_{\rm cu}$), and solar-to-hydrogen efficiency ($\eta_{\rm STH}$) calculated for SMoSe$\mid$SeWS at pH$=0$ and for SeMoS$\mid$SWSe at pH$=12.5$.}
\begin{tabular}{lrrrrr}
\toprule
Configuration & $\chi(\mathrm{O}_2)$ &$\chi(\mathrm{H}_2)$ & $\eta_{\rm abs}$ (\%) &$\eta_{\rm cu}$ (\%) & $\eta_{\rm STH}$ (\%)\\
\midrule        
SMoSe$\mid$SeWS&0.25 & 0.08  & 58.24& 29.44 & 17.14\\
SeMoS$\mid$SWSe&0.12 &   0.08& 65.39& 26.22  & 17.14\\
\bottomrule
\label{tab:STH}
\end{tabular}
\end{table}

This similar efficiency is due to a direct trade-off between $\eta_{\text{abs}}$ and  $\eta_{\text{cu}}$. Specifically, SeMoS$\mid$SWSe has a higher $\eta_{\text{abs}}=65.39\%$ compared to SMoSe$\mid$SeWS, where $\eta_{\text{abs}}=58.24\%$, because its smaller band gap allows it to harvest a broader range of the solar spectrum. However, this advantage is balanced by a lower $\eta_{\text{cu}}=26.22\%$, compared to $\eta_{\text{cu}}=29.44\%$ predicted for SMoSe$\mid$SeWS. As a result, the relative increase in absorption for one configuration is compensated by a decrease in carrier utilization in the other, leading to nearly identical overall performance for both stacking orders. These calculated values are remarkably high, comfortably surpassing the 10\% benchmark often cited for commercial viability\cite{Casandra2014,Bao22025,Ju2020,Scharber2006} and highlighting a significant potential for the application of these Janus heterobilayers as efficient and sustainable photocatalysts for water splitting.

Additional tunability can be offered by strain (homogeneously distributed between both layers in all the considered heterostructures) which is known to substantially modulate the electronic and optical properties of both conventional and Janus TMDs.\cite{Guo2020,Ramzan2025,Ramzan2023,Patel2022,XU2022} Compressive and tensile strain shift the VBM and CBM, enabling control over band offsets and gap size. For example, in the SMoSe$\mid$SWSe bilayer, where the CBM lies slightly below the $ \mathrm{H}_2 $ redox potential, suitable strain can widen the gap and align the band edges with the redox potentials required for overall water splitting, offering a promising direction for future research. As a final remark, it is important to note that under experimental conditions, ions can accumulate on the surfaces of these bilayers. Consequently, the photocatalytic performance may degrade, as the accumulated ions screen the internal electric field. Upon prolonged operation, this screening can lead to partial or even complete neutralization of the built-in field. Therefore, strategies to remove/redistribute these surface ions should be considered to maintain long-term performance by means of mechanical agitation or electrical perturbations such as the application of an external electric field.\cite{Li2014,Fu2018}

\section*{Discussion}
{The results presented so far highlight a non-trivial relationship between the internal electrostatic environment and the photocatalytic performance. Here, we rationalize these findings to provide design principles for photocatalysts based on Janus heterobilayers. Among the four considered heterostructures, only two have non-zero STH at different pH conditions, namely SMoSe|SeWS at pH = 0 and SeMoS|SWSe at pH = 12.5. Both materials are characterized by a very low potential difference ($\Delta \Phi = 0.18$~eV), and by very low dipole moments (Table~\ref{tab:electronic_properties}), due to identical chalcogen species facing each other in these configurations. Conversely, we identify a critical threshold at $\Delta \Phi \approx 1.0$~eV, beyond which the potential gradient induces a pronounced variation in the band edges. This shift either pins the CBM below the HER level or significantly contracts the fundamental band gap, rendering the system catalytically inactive despite the presence of a strong driving force. This finding suggests that the metal-to-metal chemical potential difference can be strategically used to counteract or reinforce the Janus dipole, providing a versatile and effective knob for tuning the redox ability.}

{The remarkable STH efficiency of 17.1\% achieved by the SMoSe|SeWS and SeMoS|SWSe configurations at different pH conditions is a direct consequence of this structural synergy. Different from symmetric heterobilayers, where carrier separation depends solely on the band offset, Janus heterostructures host a constructive interference between the metal-driven potential gradient and the chalcogen-driven dipoles. This dual-modulation mechanism provides an additional degree of freedom: the metal-to-metal chemical potential difference can be used to either reinforce or counteract the intrinsic Janus dipoles. By identifying the configurations where these two forces work in reciprocity, we can preserve the redox ability even under significant pH shifts, offering a significant advancement compared to apolar vdW interfaces.}

{This study delivers the electronic prerequisites that must be fulfilled by Janus TMD heterobilayers to behave as efficient photocatalysts. To achieve a comprehensive picture of the catalytic process, thermodynamic and kinetic processes must be explicitly simulated, taking into account surface-specific adsorption pathways and intermediate states under illumination. However, these investigations are secondary to the fundamental requirement of band-edge straddling. In the hierarchy of photocatalytic screening, establishing the energetic and electronic conditions for overcoming the potential barrier and separating the charge carriers represents the primary filter. By determining the specific electronic conditions and pH environments under which these requirements are met for Janus TMD heterobilayers, this work defines the necessary constraints for kinetic and thermodynamic mechanisms.}

\section*{Summary and Conclusions}
In summary, we investigated from first principles the electronic and photocatalytic properties of Janus TMD heterobilayers. All considered systems are energetically stable, {with binding energies almost one order of magnitude larger than conventional TMD heterobilayer.\cite{amin2015heterostructures,liu2023role}} Their structural properties are primarily governed by the interfacial chalcogen atoms rather than the metallic species. Except for SeMoS$\mid$SeWSe characterized by a direct band-gap at the K valley, all heterobilayers show indirect band gaps with energies within the visible light range, making them suitable for solar energy harvesting. 
{Our results identify a fundamental competition between the layer-specific Janus dipoles and the metal-to-metal chemical potential difference. We establish that a moderate built-in potential ($\Delta \Phi \approx 0.18$~eV) facilitates optimal band-edge straddling, while fields exceeding the 1.0~eV threshold lead to a breakdown of the thermodynamic conditions required for water splitting. Fulfilling these requirements,} SMoSe|SeWS bilayer is the only heterostructure suitable for overall water splitting at pH$=0$, while SeMoS|SWSe exhibits favorable characteristics at alkaline conditions (pH$\geq 12.5$). In both cases, we found a remarkably high STH efficiency of 17.14\%, {which is driven by the spatial separation of photogenerated carriers onto distinct metal surfaces. This mechanism suppresses recombination and establishes these heterostructures as high-performance electronic platforms for photocatalysis.}

In conclusion, our findings demonstrate the potential of Janus TMD heterobilayers for photocatalytic overall water splitting. {By defining the electronic descriptors that govern level alignment, this work provides a rational design to tailor vdW heterostructures with intrinsic polarization for sustainable energy production.} 
The structural and chemical versatility of the Janus TMD heterobilayers offers additional knobs to optimize the catalytic performance, such as strain, controlled defects, adsorbates, or heteroatoms.
{Furthermore, the deposition of Janus heterostructures onto switchable ferroelectric substrates represents an additional promising outlook for dynamically tunable photocatalysis.\cite{liu2025polarization,ju2023controllable,li2024intrinsically} Such architectures could enable active control over the solar-to-hydrogen conversion process, representing a compelling direction for future research in this rapidly evolving field.}

\section*{Acknowledgements}
C.C. acknowledges funding from the German Research Foundation (DFG), project number 398816777, sub-project A8. M. T. and S. S. T. thank the Research Affairs Division of the Amirkabir University of Technology (AUT).

\section*{Author contributions}
\textbf{Mostafa Torkashvand}: investigation and formal analysis,  writing original draft; \textbf{Saeedeh Sarabadani Tafreshi}:  review \&  editing; resources; \textbf{Caterina Cocchi}: conceptualization, resources, supervision, writing – review \& editing,  project administration; \textbf{Surender Kumar}: conceptualization, validation, supervision, writing original draft.

\section*{Conflicts of interest}
There are no conflicts of interest to declare.

\section*{Data availability}
Input and output files of the ab initio calculations performed in this work are available free of charge in Zenodo \href{https://doi.org/10.5281/zenodo.18226319}{DOI: 10.5281/zenodo.18226319} [record:18226319].

\balance

\bibliography{rsc} 
\bibliographystyle{rsc} 
\end{document}